\documentclass{iau} 
\usepackage{graphicx}

\title[IAUS 346.~~High-mass X-ray binaries] %% give here short title %%
{Massive star winds and HMXB donors}
\author[Andreas A.C. Sander]   %% give here short author list %%
{Andreas A.C. Sander$^{1,2}$
}

\affiliation{$^1$Armagh Observatory and Planetarium, \\
College Hill, Armagh BT61 9DG, Northern Ireland, UK \\ 
email: {\tt Andreas.Sander@armagh.ac.uk} \\[\affilskip]
$^2$Institut f{\"u}r Physik und Astronomie, Universit{\"a}t Potsdam, \\
Karl-Liebknecht-Str. 24/25, D-14476 Potsdam, Germany}

\pubyear{2018}
\volume{346}  %% insert here IAU Symposium No.
\jname{High-mass X-ray binaries: illuminating the passage from massive binaries to merging compact objects}
\editors{L.M.Oskinova, E. Bozzo, D. Gies, \& D. Holz, eds.}
\begin{document}

\maketitle

\begin{abstract}
Understanding the complex behavior of High Mass X-ray binaries (HMXBs) is 
not possible without detailed information about their donor stars. While crucial,
this turns out to be a challenge on multiple fronts. First, multi-wavelength 
spectroscopy is vital. As such systems can be highly absorbed, this is often
already hard to accomplish. Secondly, even if the spectroscopic data is available,
the determination of reliable stellar parameters requires sophisticated model
atmospheres that accurately describe the outermost layers and the wind of the donor star.

For early-type donors, the stellar wind is radiatively driven and there is a
smooth transition between the outermost layers of the star and the wind. The intricate 
non-LTE conditions in the winds of hot stars complicate the situation even
further, as proper model atmospheres need to account for a multitude of physics
to accurately provide stellar and wind parameters. The latter are especially crucial 
for the so-called ``wind-fed'' HXMBs, where the captured wind of the supergiant donor is 
the only source for the material accreted by the compact object.

In this review I will briefly address the different approaches for treating stellar winds
in the analysis of HMXBs. The fundamentals of stellar atmosphere modeling are discussed,
also addressing the limitations of modern models and examples from recent analysis results for 
particular HMXBs. Furthermore, the path for the next generation of stellar 
atmosphere models will be outlined, where models can be used not only for measurement 
purposes, but also to make predictions and provide a laboratory for theoretical conclusions. 
Stellar atmospheres are a key tool in understanding HMXBs, e.g. by providing insights about
the accretion of stellar winds onto the compact object, or by placing the studied systems 
in the correct evolutionary context in order to identify potential gravitational wave (GW) progenitors.

\end{abstract}

\firstsection % if your document starts with a section,
\section{Introduction}

High-mass X-ray binaries (HMXBs) consist of a compact object - either a neutron star or a black hole - accreting
material from a companion. This companion is a massive star with initially more than about $8$ times the mass 
of our sun. HMXBs can be classified into three different types of systems:
\begin{enumerate}
  \item \textbf{Wind-fed} systems, where the compact object accretes material by capturing a fraction the donor star's wind
	\item \textbf{Roche-lobe overflow} (RLOF) systems, where the material from the donor is (mainly) accreted via the mass-loss through the 
	              inner Lagrangian point (L$_1$) 
	\item \textbf{Be X-ray binaries} (BeXRBs) where the donor is a Be-star with a decretion disk and the compact object is (mainly) fed by accreting disk material
\end{enumerate}

Although not inherent to their definition, the donor stars usually found in these systems are all hot and luminous stars, having $T_\mathrm{eff} > 10\,$kK 
and thus a strong UV flux. This makes HMXBs excellent tracers of star formation and young populations in galaxies. The spectral types of the
massive donor stars are typically early B or late O-type stars, but also earlier O-subtypes and even some Wolf-Rayet (WR) stars are known as donors.

While BeXRB systems are quite distinct, the border between wind-fed and RLOF systems is not that obvious, especially if the radii of the donor stars are not well
constrained and thus it can be unclear whether a donor really fills its Roche lobe or not. Moreover, Bondi-Hoyle-Lyttleton (BHL) 
accretion mechanism \cite[(Hoyle \& Lyttleton 1939; Bondi \& Hoyle 1944; Davidson \& Ostriker 1973)]{HL1939,BH1944,DO1973} yields that the possible X-ray luminosity
due to accretion $L_\mathrm{X}$ crucially depends on the wind parameters of the donor star
\begin{equation}
	L_\mathrm{X} \propto \frac{\dot{M}_\mathrm{donor}}{\left[v_\mathrm{orb}(d)^2+v_\mathrm{wind}(d)^2\right]^{3/2} \cdot v_\mathrm{wind}(d)} \approx \frac{\dot{M}_\mathrm{donor}}{v_\mathrm{wind}^4(d)}\mathrm{,}
\end{equation}
in particular the wind speed $v_\mathrm{wind}$ at the orbital separation $d$ of the compact object. Sophisticated spectral analyzes of the donor stars
using state-of-the-art model atmospheres to determine the stellar and wind parameters of the donor stars in HMXBs are therefore a pivotal instrument for
a better understanding of this important evolutionary stage towards compact object binaries.

\section{Modeling the winds in HMXBs}
  \label{sec:model}

To study the impact of winds in HMXB systems, two sorts of approaches have been applied, both having their advantages and disadvantages. On the one side,
the situation can be studied with the help of hydrodynamics codes, allowing potentially for a multidimensional and time-dependent treatment in a rather
complex geometry. However, the detailed treatment in terms of space and resolution comes at the cost of simplified wind physics. Two typical examples are
the radiative transfer, which is usually approximated, and the use of a so-called ``ionization parameter'', which approximates the calculation
of a detailed ionization balance as a single number entering the actual calculations. Seminal work in this field has been done by \cite[Blondin et al. (1990)]{Blondin+1990},
followed up nowadays by the works of e.g. \cite{Manousakis+Walter2015} or \cite[El Mellah et al. (2018a,b)]{ElMellah+2018a,ElMellah+2018b}. 

A second, rather different approach is the use of stellar atmospheres. Well established in the field of isolated massive stars, they provide a detailed 
physical treatment for the donor, including a full non-LTE calculation of the ionization structure and a radiative transfer performed in the co-moving frame.
Here, the costs are a simplified geometry and usually also a stationary wind. Traditionally developed for spherically symmetric situations, the compact object
in this approach is rather treated as a ``disturbance'' instead of being fundamentally taken into account from first principles. Applications to HMXBs are quite
limited in terms of the number of systems studied, but have been done for quite a while, ranging from classic spectroscopic stellar 
analysis \cite[(e.g. Clark et al. 2002)]{Clark+2002} to studies of the effect of X-ray irradiation on the radiative
driving of the donor star  \cite[(e.g. Krti{\v c}ka et al. 2012, Sander et al. 2018)]{Krticka+2012,Sander+2018}. Essentially, the two approaches complement each 
other. To fully understand the role of the donor stars in HMXBs, a proper atmosphere analysis is essential and thus the second approach will be the focus of this article.

Hot, massive stars have inherent mass outflows, called ``stellar winds''. These winds can reach up to several thousand km\,s$^{-1}$ with typical mass-loss rates
on the order of $10^{-6}\,M_\odot$\,yr$^{-1}$ for O supergiants at solar metallicity. The wind is accelerated by momentum transfer from stellar continuum 
photons to metal ions, which happens mainly due to absorption in spectral lines. Due to the fact that
the re-emission is isotropic, while the absorbed radiation is radially coming from the star, there is a net momentum away from the stellar surface. For O and B stars, the most
prominent spectral signatures of stellar winds are the P Cygni profiles found in the ultraviolet resonance lines of e.g. C\,\textsc{iv}. For more dense winds,
some optical line profiles such as H$\alpha$ will turn from an absorption into an emission. For sufficiently dense winds, such as found in Galactic WR stars, the
whole spectrum is formed in the wind and all optical lines are in emission. However, the lack of optical emission lines on the other hand does not imply the
absence of a considerable wind.

The theoretical description of these radiation-driven winds was pioneered by \cite[Lucy \& Solomon (1970)]{LS1970} and \cite[Castor, Abbott, \& Klein (1975)]{CAK1975}.
The initials of the latter provided the name for the so-called ``CAK theory''. 
While considerably improved in the 1980s \cite[(e.g. Friend \& Abbott 1986, Pauldrach et al. 1986)]{FA1986,Pauldrach+1986}, the fundamental
premises of CAK such as the assumption that all acceleration is provided either by free electron or line scattering remain until today. A major outcome of the theory is the
so-called ``$\beta$-law'' describing the radial behavior of the wind velocity
\begin{equation}
  v(r) = v_\infty \left( 1 - \frac{R_\star}{r} \right)^\beta
\end{equation}
for a massive star with the help of a parameter $\beta$. This essentially analytic description of $v(r)$ is widely used, allowing an easy
determination of the velocity or density of the stellar wind in more complex calculations or simulations. Nonetheless, there are various cases where the
CAK formalism fails to correctly describe the observed stellar winds, especially but not limited to the situations where winds get more dense and the average 
photon leaving the star is scattered more than once (``multiple scattering''). It is thus favorable to go beyond these limitations in a sophisticated modeling 
of a stellar atmosphere. Multiple challenges have to be accounted for in this task, including
%Basically, the modeling of a hot star atmosphere with a line-driven wind requires to deal with multiple challenges, including

\begin{itemize}
  \item the intricate non-LTE situation,
	\item a radiative transfer in an expanding atmosphere accounting for multiple scattering,
	\item sufficient model atoms accounting for all elements that are either detectable in the wind or influence the stratification.
\end{itemize}

The non-LTE situation implies that we cannot rely on the Saha-Boltzmann statistics, but instead assume a statistical equilibrium, i.e.
a balance of total gain and loss rates for all considered atomic levels. Unlike in cool stars, it is not sufficient to determine just
the ionization balance. Instead we need to obtain the population numbers for a considerable amount of levels in each relevant ionization 
stage of each element taken into account. Even when considering only a handful of elements, this quickly leads to a system of a few hundreds 
of rate equations that needs to be solved.

Obtaining a proper model atom is especially tricky for the elements of the iron group which have thousands of 
levels and millions of line transitions. This introduces a so-called ``blanketing'' effect, significantly altering the atmosphere stratification.
An explicit treatment of all these levels is impossible. Thus, basically all modern atmosphere codes use a concept
going back to \cite[Anderson (1989)]{Anderson1989} and \cite[Dreizler \& Werner (1993)]{DW1993}
where levels in a certain energy range are grouped into a ``superlevel''. While the superlevels are then treated
in full non-LTE, the levels inside a superlevel are assumed to follow the Boltzmann statistics. 

In an expanding, non-LTE environment, the determination of the electron temperature stratification is also not
trivial. Going back to the ideas of \cite[Uns{\"o}ld (1951, 1955)]{U1951,U1955Book}
and \cite[Lucy (1964)]{L1964}, temperature corrections can be obtained from the equation of radiative
equilibrium and it's integral describing the conservation of the total flux. Alternatively, one can also
obtain the corrections from calculating the electron thermal balance going back to \cite[Hummer \& Seaton (1963)]{HS1963}.
Since both of these two -- or three, depending on how one counts -- methods have their strengths and
weaknesses, some stellar atmosphere codes can also use a combination of them.

The radiative transfer is performed in the co-moving frame (CMF). While this is computationally considerably more expensive than the CAK formalism, this rigorous approach
considers all contributions to the radiative acceleration and implicitly accounts for various effects such as multiple scattering.
Combining the previously sketched techniques is a task of its own, as the problems are highly coupled. The CMF radiative transfer introduces a coupling in space, 
and the solution of the statistical equations comes with an intrinsic coupling in frequency, thus we are faced with a problem coupled in both, space and frequency.
The solution is an accelerated lambda iteration \cite[(ALI, e.g. Hamann 1985)]{H1985} until a consistent solution of the calculations (i.e. solution of the 
statistical equations, radiative transfer, temperature corrections) is reached. Beside this, one more layer
of complexity is added by the requirement to account for both the quasi-hydrostatic layers of the star as well as the supersonic wind on top. While for an analysis 
purpose it is often sufficient to assume a $\beta$-type velocity law in the wind, the proper quasi-hydrostatic treatment requires additional updates of the density
stratification \cite[(see, e.g., Sander et al. 2015)]{Sander+2015}. 

When a converged atmosphere stratification is reached, the emergent spectrum is calculated in a final radiative transfer calculation, now performed in the observer's frame.
The resulting synthetic spectrum can then be compared to observations in order to check whether a model with certain assumed parameters is a sufficient representation of the star. 
If matching, the complexity of the model atmospheres then becomes a huge advantage: Assuming sufficient observed spectra are available, preferably in multiple wavelength regimes, 
stellar atmosphere models provide a multitude of information about the donor star, including the stellar and 
wind parameters (e.g. $T_\mathrm{eff}$, $\log g$, $L$, $v_\infty$, $\dot{M}$, \dots), the chemical abundances, 
and also the mechanical and ionizing feedback. This level of knowledge is essential for a proper understanding of the donor and thus the HMXB system in general.

\section{The parameters of HMXB donors}

Given the progress in analyzing single massive stars, it is tempting to apply the insights gained there to an HMXB donor star 
of the same spectral type. However, donor stars are not necessarily identical to massive single stars. Firstly, the compact object next to them might be small 
in size, but not in influence. The breaking of the spherical symmetry and the influence of the X-rays irradiating the donor star can make an important difference. 
Secondly, the donor star in an HMXB system has a very distinctive history. The existence of the compact
object points to the fact that the current donor has been the secondary in a former binary system and thus was likely a mass gainer during an earlier stage of mass transfer
\cite[(e.g. Vanbeveren \& De Loore 1994)]{VdL1994}. 
From the observational and analytical point of view, various studies have confirmed that donor stars are more luminous than single stars of the same mass \cite[(e.g. Conti 1978,
Vanbeveren et al. 1993)]{C1978,Vanbeveren+1993}. \cite{K2001} summarized that for their luminosity, the donor stars have lower masses and radii compared to
normal OB supergiants. While the total number of systems studied in detail so far is too low to give final answers, these results nonetheless outline the peculiarity of
HMXB donor stars, thus making it problematic to simply adopt parameters from isolated stars. Still this is and sometimes even has to be done as HMXB systems are often well 
studied in X-rays, but poorly at other wavelength regimes. This discrepancy of knowledge has two major reasons of very different origin: The observational reason is the high 
interstellar absorption for many of the systems which prevents their observation in the UV and often even in the optical. With the UV being the most important wind diagnostic regime for O and B-type stars, this leads to a significant limitation of statistics as only a handful of systems can actually be studied in this regime.
The second, more sociological reason is the interdisciplinary nature of HMXB systems, combining stellar winds with accretion mechanisms and high-energy physics. A proper understanding requires experts from various fields and when new systems are discovered in X-rays, there is a delay until observations in other wavelength regimes have been made and follow-up science is underway.
Thus, it is more important than ever to bring experts from different fields together early-on in the process.
%X-wind mention here`?

%(Mart´ınez-N´u˜nez et al. 2017) lists 31 wind-fed
%systems:
%I 28 of them have a spectral type designation
%I about 7 of them are analyzed with atmosphere models
	
The analysis of the donor star with different modern stellar atmospheres has been done for about a handful of systems. While differing in their computational details, sometimes quite significantly, three of them, namely PoWR \cite[(Hamann \& Schmutz 1987, Hamann \& Koesterke 1998, Gr{\"a}fener et al. 2002)]{HS1987,HK1998,GKH2002}, CMFGEN \cite[(Hillier 1990, Hillier \& Miller 1998)]{H1990,HM1998}, and FASTWIND \cite[(Santolaya-Rey et al. 1997, Puls et al. 2005)]{Santolaya-Rey+1997,Puls+2005}, are state-of-the-art tools for analyzing hot stars with expanding atmospheres. PoWR and CMFGEN exactly
reflect the scheme sketched in Sect.\,\ref{sec:model}, while FASTWIND uses an approximate blanketing approach, getting lower computation times for their models at the cost of a smaller range of applicable stars, e.g. excluding Wolf-Rayet stars. For purely hydrostatic atmospheres, also the TLUSTY code \cite[(Hubeny 1988, Hubeny \& Lanz 1995)]{H1988,HL1995} is an option. However, since in most HMXB systems -- maybe aside from the BeXRBs -- the wind is not negligible, a model atmosphere inherently accounting for the wind is highly recommended as otherwise the deduced stellar parameters could be wrong due to wind emission filling up the absorption profiles.

% \cite[(e.g. Fraser et al. 2010)]{Fraser+2010}

The METUJE code by \cite{Krticka+2012,Krticka+2015} takes a slightly different approach by focussing on the calculation of mass-loss rates instead of reproducing observed spectra. This concept will be discussed further in Sect.\,\ref{sec:hydro}. There are more atmosphere codes on a comparable level of sophistication, such as WM-basic \cite[(Pauldrach 1987, Pauldrach et al. 1994, 2001)]{Pauldrach1987, Pauldrach+1994,Pauldrach+2001} or PHOENIX \cite[(Hauschildt 1992; Hauschildt \& Baron 1999, 2004)]{Hauschildt1992,HB1999,HB2004}, but both have shifted their focus to other areas (supernovae for WM-basic, cooler stars for PHOENIX) and thus to our knowledge they have so far not been applied to HMXB donors.  

%
%Codes:
%
%PoWR: e.g. Vela X-1, IGR J17544-2619 (Gim´enez-Garc´ıa et al. 2016),
%X1908+075 (Mart´ınez-N´u˜nez et al. 2015)
%I CMFGEN: e.g. 4U1700-37 (Clark et al. 2002), 4U 1907+09 (Cox et al.
%2005), GX301-2 (Kaper et al. 2016)
%I FASTWIND: e.g. IGR J16465-4507 (Chaty et al. 2016)
%I METUJE: more theory-focussed, ˙M calculations (e.g. Krtiˇcka et al.
%2012, 2015)
%I TLUSTY: e.g. Vela X-1 (Fraser et al. 2010), but static treatment (i.e.
%no wind), thus fully valid only for hot donors with negligible winds
%Further codes in the field:
%I WM-basic: UV-focused, SN treatment, pressure broadening
%I PHOENIX: typically	
	
\begin{table}
  \begin{center}
  \caption{Examples of wind-fed system analyses}
  \label{tab:params}
 {\scriptsize

	\begin{tabular}{|l|c|c|c|c|c|c|c|l|}\hline
    \textbf{System} & \textbf{sp.~Type}  & $T_\mathrm{eff}$ & $\log g$ &  $X_\mathrm{He}^\mathrm{a}$ &  $X_\mathrm{N}^\mathrm{a}$ & $\log \dot{M}$         &  $v_\infty$      &   \textbf{Reference}  \\
	                  &           &   [kK]           &   [cgs]  &    &  [$10^{-3}$]  & [$M_\odot\,$yr$^{-1}$] &  [km\,$^{-1}$]   &    \\  \hline
 		GX 301-2        & B1I       & $19$  &    $2.38$   &   $0.52$  &  $5.25$ & $-5.0$  & $305$ & {\tiny  \cite{Kaper+2006}}  \\
		4U 1909+07      & B0I..B3I & $23$   &    $3.2$    &   $0.26^\mathrm{b}$  &  $0.6^\mathrm{b}$   & $-6.5$  & $500$ & {\tiny  \cite{MartinezNunez+2015}}  \\
	  Vela X-1        & B0.5I    & $26$   &    $2.86$   &   $0.34$  &  $1.8$ & $-6.2$  & $700$ & {\tiny  \cite{GimenezGarcia+2016}} \\
		IGR J16465-4507 & B0.5I    & $26$   &    $3.10$   &   $0.5$   &  $5.5$  &  n/a    &  n/a  & {\tiny  \cite{Chaty+2016}}  \\
		IGR J17544-2619 & O9.5I    & $29$   &    $3.25$   &   $0.37$  &  $2.2$ & $-5.8$ &  $1500$ & {\tiny \cite{GimenezGarcia+2016}}  \\
    4U 1907+09      & O8I..O9I & $30$   &    $3.1$    &   $0.26^\mathrm{b}$  & $0.6^\mathrm{b}$  & $-5.1$ & $1700$  & {\tiny \cite{Cox+2005}} \\
		4U 1700-37      & O6.5I    & $35$   &    $3.5$    &   $0.44$  & $10$  & $-5.0$ & $1750$ & {\tiny \cite{Clark+2002}} \\
 	\hline
	\end{tabular}
%
  %\begin{tabular}{|l|c|c|c|c|}\hline 
%{\bf Mineral} & {\bf Size [$\mu$m]} & {\bf Isotopic Signatures} & {\bf Stellar} & {\bf Contri-} \\ 
   %&  {\bf abund.}  [ppm]$^1$ & & {\bf Sources} & {\bf bution$^2$} \\ \hline
%diamond & $~0.0026$ & Kr-H, Xe-HL, Te-H & supernovae & ? \\
   %& ~1500 & & & \\ \hline
%silicon & $~0.1-10$ & enhanced $^{13}$C, $^{14}$N, $^{22}$Ne, s-process elem. & AGB stars & $> 90$~\% \\
%carbide & $~30$ & low $^{12}$C/$^{13}$C, often enh.\ $^{15}$N & J-type C-stars (?) & $< 5$~\% \\
 %& & enhanced $^{12}$C, $^{15}$N, $^{28}$Si; extinct $^{26}$Al, $^{44}$Ti & Supernovae & 1~\% \\
 %& & low $^{12}$C/$^{13}$C, low $^{14}$N/$^{15}$N & novae &  $0.1$~\% \\ \hline
%graphite & $~0.1-10$ & enh.\ $^{12}$C, $^{15}$N, $^{28}$Si; extinct $^{26}$Al, $^{41}$Ca, $^{44}$Ti &
 %SN (WR?) & $< 80$~\% \\ 
 %& $~10$ & s-process elements & AGB stars & $> 10$~\% \\
 %& & low $^{12}$C/$^{13}$C  & J-type C-stars (?) & $< 10$~\% \\
 %& & low $^{12}$C/$^{13}$C; Ne-E(L) & novae & 2~\% \\ \hline
 %corundum/ & $~0.1-5$ & enhanced $^{17}$O, moderately depl. $^{18}$O & RGB / AGB & $> 70$~\% \\
 %spinel/ & $~50$ & enhanced $^{17}$O, strongly depl. $^{18}$O & AGB stars & 20~\% \\
 %hibonite & & enhanced $^{16}$O & supernovae & 1~\% \\ \hline
 %silicates & $~0.1-1$ & similar to oxides above & & \\
 %&  $~140$ & & & \\ \hline
 %silicon & $~1$ & enhanced $^{12}$C, $^{15}$N, $^{28}$Si; extinct $^{26}$Al & supernovae & 100~\% \\
 %nitride & $~ 0.002$ & & & \\ \hline
  %\end{tabular}
  }
 \end{center}
\vspace{1mm}
 \scriptsize{
 {\it Notes:}\\
   $^\mathrm{a}$mass fraction
   $^\mathrm{b}$assumed solar abundance
%  $^1$For the abund.\ (in wt.\ ppm) the reported maximum values from different meteorites are given. \\
%  $^2$Note uncertainty about actual fraction of diamonds that are pre-solar and for fraction of graphite attributed to SN and AGB stars (see discussion in text).
 }%
\end{table}

A selection of wind-fed HMXB systems analyzed with current state-of-the-art atmosphere codes is compiled in Table\,\ref{tab:params}. Essentially, all of these systems have late O- or
early B-type donor stars. In systems where abundances could be determined, the helium and nitrogen fractions are enhanced, clearly indicating that they are evolved stars. The wind parameters
of the various donors differ quite significantly, even if their spectral type is rather similar. The terminal velocities show a trend with increasing $T_\mathrm{eff}$, but the mass-loss
rates do not reflect this. Instead we see a signficantly higher mass-loss rate for the coolest star in the sample compared to those with slightly higher temperatures. While there are
effects due to the compact objects and its accretion as we will discuss below, this significant change is most likely not a result of this, but of the so-called bi-stability jump. 
This phenomenon is not exclusive to HMXB donors, but a general effect seen in the regime of early B supergiants in a temperature regime of $T_\mathrm{eff} \approx 20...25\,$kK.
Found theoretically by \cite{PP1990} and confirmed observationally by \cite{Lamers+1995}, a drastic increase in the mass-loss rate $\dot{M}$ is seen compared to stars with higher
$T_\mathrm{eff}$, accompanied by a considerable decrease of the terminal wind velocity $v_\infty$. Studied since its discovery \cite[(e.g. by Vink et al. 1999, Petrov et al. 2016,
Keszthelyi et al. 2017)]{Vink+1999,Petrov+2016,Keszthelyi+2017}, the bi-stability jump is attributed to a change in the ionization balance - mainly of iron - but a lot of details still remain unclear, including the
precise jump conditions and the metallicity dependence. 

In HMXBs, X-ray irradiation might further influence the ionization structure, thus further complicating
the situation. \cite{HM1977} noticed that since the irradiation is coming from the orbiting compact object, the effects should be seen in the phase-dependent profiles of UV resonance lines, but in fact this
so-called ``Hatchett-McCray effect'' is not observed for all systems. Approximated radiative transfer models by \cite{vanLoon+2001} revealed that the precise orbital configuration can have a significant influence on the appearance of this effect and demonstrated that for systems like 4U\,1700$-$37 indeed no significant phase-dependent profile change should occur.

A significant change of the ionization balance also affects the stellar wind. A considerable X-ray irradiation effectively harms the wind in the direction towards the compact object. The depopulation of important ionization stages leads to the effect that the wind is no longer accelerated and might even slow down \cite[(Krti{\v c}ka et al. 2012)]{Krticka+2012}. However,
for the mass-loss rate itself to be changed, almost the whole donor atmosphere down to the stellar surface would need to be affected, which is probably not the case in most systems \cite[(Sander et al. 2018)]{Sander+2018}. 

The X-ray flux variabilities seen in HMXBs further suggest that the winds of the donor stars are not homogeneous \cite[(e.g. in’t Zand 2005, van der Meer 2005, Negueruela et al. 2008, Oskinova et al. 2012)]{iZ2005,vdM+2005,Negueruela+2008,Oskinova+2012}.
Indeed, smooth winds are not expected even from single stars, where small scale structures, often referred to as ``clumping'' are a part of the standard picture. However, the theoretical concepts
for clumping in single star winds predict rather small clumps, which do not really match the larger clump sized inferred from the X-ray observations so far. The origin of clumping itself
is still a matter of debate, with the classic idea of the line-driven or line deshadowing instability \cite[(LDI, e.g. Owocki et al. 1988, Feldmeier et al. 1997, Dessart \& Owocki 2003, Sundqvist et al. 2018)]{Owocki+1988,Feldmeier+1997,DO2003,Sundqvist+2018}
facing certain challenges, e.g. regarding the predicted onset of the clumps. Envelope instability or sub-surface convection \cite[(e.g. Glatzel 2008, Cantiello et al. 2009, 
Jiang et al. 2015)]{Glatzel2008,Cantiello+2009,Jiang+2015} has been suggested as an alternative that would allow for a deeper clumping onset around opacity peaks, e.g. of iron in main sequence stars. Small-scale clumping also likely co-exists with large scale structures, attributed to co-rorating interaction regions (CIRs) spectroscopically manifesting in so-called discrete absorption components (DACs).
A detailed discussion about clumping and winds in HMXBs was recently published in an extensive review by \cite{HMXBReview2017}.

\section{From measurements to laboratories}
  \label{sec:hydro}

Traditionally, stellar atmosphere models have been used to measure the stellar and wind parameters.
However, their complexity also allows us to use them as virtual laboratories, where we can test what for example the
imprint of a certain change of parameters is on the stellar spectrum. Moreover, we can predict the wind parameters ($v_\mathrm{wind}(r), \dot{M}_\mathrm{donor}$)
instead of measuring them by including the solution of the hydrodynamic (HD) equation of motion
\begin{equation}
  \label{eq:hydro}
     v \left( 1 - \frac{a_\mathrm{s}^2}{v^2} \right) \frac{\mathrm{d} v}{\mathrm{d} r} = a_\mathrm{rad}(r) - g(r) + 2 \frac{a_\mathrm{s}^2}{r} - \frac{\mathrm{d} a_\mathrm{s}^2}{\mathrm{d} r}\mathrm{,}
\end{equation}
which has to be fulfilled at all depths in a self-consistent atmosphere model. Due to the fact that the
hydrodynamic Eq.\,(\ref{eq:hydro}) has a critical point, the additional constraint of requiring a smooth
transition of $v(r) \equiv v_\mathrm{wind}(r)$ through this point is introduced. This can be translated into a condition for the 
mass-loss rate $\dot{M}_\mathrm{donor}$. Thus, a converged model that also fulfills Eq.\,(\ref{eq:hydro}), essentially predicts the 
fundamental wind parameters from a given set of stellar parameters.

While this concept might sound rather simple and goes back to \cite{LS1970}, its actual implementation is not.
Early efforts were made by \cite{Pauldrach+1986}, using a pure CMF line force implementation. WM-basic \cite{Pauldrach+2001} later
used the concept together with a Sobolev-based approach. Predicting mass-loss rates in this way is also the heart of
the more theory-focussed METUJE code. In \cite{KK2004}, a Sobolev approach was used, but since \cite{KK2010} also a 
CMF radiative transfer is standard. \cite{GH2005} performed the first complete implementation into a CMF-based analysis code,
namely PoWR. Successfull applications for a WC and later also a grid of WN models \cite[(Gr{\"a}fener \& Hamann 2008)]{GH2008}
were performed using an implementation based on a generalized force multiplier concept. To extend this method to further regimes,
a new implementation with a different technique was added to PoWR by \cite{Sander+2017}, yielding also hydrodynamically consistent 
models for O and B stars. Recently, \cite[Sundqvist \& Puls (2018)]{SP2018} announced the solution of the hydrodynamic equation of motion also as 
one of their goals for a future FASTWIND update.

\begin{figure}[t]
\begin{center}
 \includegraphics[width=0.99\textwidth]{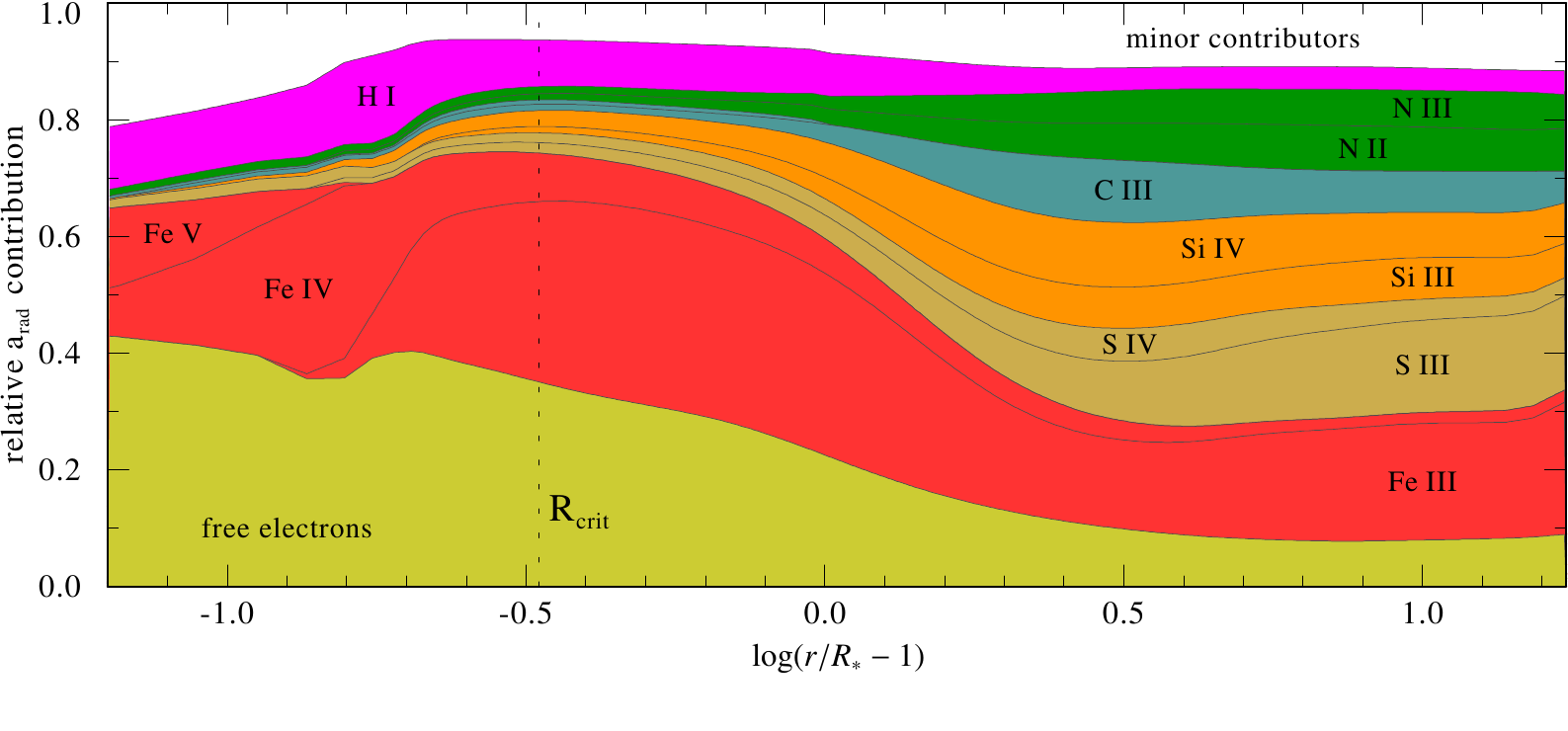} 
 \caption{Relative contributions to the radiative acceleration in the donor star of Vela X-1}
   \label{fig:acccontrib}
\end{center}
\end{figure}

Utilizing the new PoWR implementation, \cite{Sander+2018} published a self-consistent model for the donor star of Vela X-1.
This included a standard model as well as additional models with different levels of X-ray irradiation in order to study the irradition
effect on the radiative driving and the resulting wind velocity field. The study revealed that in the donor of Vela X-1, 
apart from free electrons, Fe\,\textsc{iii} is the leading ion driving the wind, followed by S\,\textsc{iii} (see Fig.\,\ref{fig:acccontrib}). 
However, just because Fe\,\textsc{iii} is
the leading driver, this does not imply the Fe\,\textsc{iii} is the main Fe ionization stage in the wind. In fact, most of
the material is in Fe\,\textsc{iv}, but its driving contribution is just not important in the outer wind. The irradiation of
X-rays from the accretion onto the neutron star can be large enough to shut down the further acceleration in the affected cone and
thus reduce the terminal wind velocity there compared to an isolated massive star, but they do not penetrate deep enough to reach
the layers where the mass-loss rate is set, thus leaving it basically unaffected compared to the unperturbed situation.

An important detail revealed by the model in \cite{Sander+2018} is the tailored wind velocity law and its deviation from
a typical $\beta$-law in the wind onset and inner wind region. This is crucial as the neutron star in Vela X-1 is sitting at
a distance of about $1.8\,R_\ast$, a rather typical value for wind-fed HMXBs, and it is exactly this region where the velocity
is about a factor of two lower than one would infer from using a $\beta$-law. Since Bondi-Hoyle accretion approximately yields
$L_\mathrm{X} \propto \dot{M}/v^4$, the potential $L_\mathrm{X}$ is very sensitive to such deviations and a velocity that is 
by a factor of two lower could already power almost an order of magnitude more X-ray luminosity.

Hydrodynamically consistent models are a significant step forward, but of course still one-dimensional and thus also they
have their limitations. A long-term thus would be to merge such a sophisticated treatment of the radiative transfer with
multi-dimensional hydrodynamic simulations, but this is beyond current computational limits.

\section{Summary and conclusions}

The donor stars of high-mass X-ray binary systems are -- with the exception of the BeXRBs -- typically late O- or early B-type supergiants. The details
of their wind parameters can differ significantly due to the bi-stability regime and further the different orbital configurations. In many cases,
the donor properties are rather poorly constrained. Photometry is usually not sufficient to give a valid characterization of the donor. Spectra are
required to avoid inferring wrong quantities or evolutionary scenarios.
To derive the stellar wind parameters of the donor stars is a difficult task, especially as donor stars are peculiar in their properties when
compared to isolated massive stars of the same spectral type. An atmosphere analysis with a state-of-the-art model atmosphere code is strongly 
recommended to get a reliable handle on the wind parameters. This is even more important if no UV spectra of the donor are available, despite
the fact that also the atmosphere analysis might be limited with regards to some parameters.
The non-homogeneous nature (or ``clumpyness'') of stellar winds are supported both from the single star observations as well as the accretion 
X-ray variability. Clumps might be able to explain a variety of observed phenomena, but also add a considerable layer of complexity to the models.
The presence of a strong X-ray source near the donor star has a noticeable effect on its (outer) atmosphere and can change the wind ionization
and acceleration. As the causing X-ray illumination is phase-dependent, some systems also show phase-dependent changes of the spectrum, in
particular in the UV resonance lines.
Beside the detailed, but one-dimensional stellar atmosphere models, another important tool to tackle the complex behaviour of donor winds in HMXB 
systems as multi-dimensional, hydrodynamic models, having a kind of complementary approach with their complex geometry, but simplified wind physics. 
So far, we are not in a stage to combine both approaches, but first steps, where results from one approach are incorporated in the other one in
a parametrized form are on the way.

%\input{bibdefinitions.tex}
%\bibliographystyle{aa} % style aa.bst
%\bibliography{galwcgaia}

\begin{discussion}

\discuss{L.~Kaper}{The X-ray eclipse presents a way to measure the radius of the OB supergiant with
high accuracy. This parameter is often poorly constrained by stellar atmosphere models. Do you use these
radius determination?}

\discuss{A.A.C.~Sander}{This is a good point. We do not explicitly use this in the model for the Vela X-1 donor,
 since the stellar parameters are taken from \cite{GimenezGarcia+2016}, but as far as I am aware
 their analysis makes use of this. Moreover, we did use radius constraints in the analysis of the X1908+075 donor
 where we went with the rather unconventional way to fix $T_\ast$ and $R_\ast$ to constrain the luminosity 
 and distance.}

\discuss{E.P.J.~van den Heuvel}{In some of the highly obscured Integral supergiant HMXBs, the donor is a later O supergiant,
but from the high obscuration you would expect the wind to be much slower and denser than in a single O supergiant. Has
any work been done on how such winds are slowed down by the X-ray emission?}

\discuss{A.A.C.~Sander}{As far as I am aware there is no explicit work with respect to these highly obscured Integral sources. However,
depending on the orbital situation, a significant shut down of the wind acceleration and thus low velocities due to X-ray irradiation
seems quite reasonable. This effect has been studied in \cite{Krticka+2012} and also \cite{Sander+2018}, but both times based on the
Vela X-1 system. J.~Krti{\v c}ka will talk much more about the effect of X-ray irradiation (see these proceedings).}

\discuss{A.~Kashi}{Do you assume clumping in your simulations or do you get the clumping factor as a result?}

\discuss{A.A.C.~Sander}{A depth-dependent clumping stratification is assumed in the Vela X-1 donor model. We account 
for the so-called ``microclumping'', i.e. optically thin clumps, which is the standard approach in current stellar atmosphere models. For
a detailed optically thick clumping a two-component calculation would be required, which is computationally expensive and requires
a considerable extension of a model code. Using a Monte Carlo approach, detailed calculations have been performed by \cite[{\v S}urlan et al. (2012, 2013)]{Surlan+2012,Surlan+2013}.
For standard model atmospheres, an approximate treatment of optically thick clumping that does not rely on a two-component calculation has 
recently be introduced to FASTWIND by \cite{SP2018}.}

\discuss{D.~Gies}{Are the ionization calculations for the on-axis position in the binary?}

\discuss{A.A.C.~Sander}{Yes. Since the atmosphere model is essentially 1D, the performed calculations would best represent the on-axis situation.}

\discuss{F.~Mirabel}{Feedback from compact HMXBs is in the form of X-rays and jets (e.g. Cyg X-1, SS\,433), and we know
that the mechanical energy from the jets may be larger than that by X-rays. This feedback from jets complicate further the modeling
of accretion winds.}

\discuss{A.A.C.~Sander}{Fair point. While the jets usually should not directly hit the surface of the donor star and modify the atmosphere, their existence
would complicate the proper interpretation of the observations and measurements. Fortunately the Vela X-1 system is not known for having jets.}

\end{discussion}

\end{document}